\documentclass[twoside,11pt]{article}

\usepackage{blindtext}
\usepackage{breakcites}
\usepackage{parcolumns}
\usepackage{listings}
\usepackage{xcolor}
\usepackage{todonotes}
\usepackage{breakurl}

\definecolor{codegreen}{rgb}{0,0.6,0}
\definecolor{codegray}{rgb}{0.5,0.5,0.5}
\definecolor{codepurple}{rgb}{0.58,0,0.82}
\definecolor{backcolour}{rgb}{0.95,0.95,0.92}

\lstdefinestyle{mystyle}{
    backgroundcolor=\color{backcolour},   
    commentstyle=\color{codegreen},
    keywordstyle=\color{magenta},
    numberstyle=\tiny\color{codegray},
    stringstyle=\color{codepurple},
    basicstyle=\ttfamily\footnotesize,
    breakatwhitespace=false,         
    breaklines=true,                 
    captionpos=b,                    
    keepspaces=true,                 
    numbers=left,                    
    numbersep=5pt,                  
    showspaces=false,                
    showstringspaces=false,
    showtabs=false,                  
    tabsize=2
}

\usepackage[preprint,hyperref]{jmlr2e}

\usepackage{lastpage}
\jmlrheading{23}{2024}{1-\pageref{LastPage}}{1/21; Revised 5/22}{9/22}{21-0000}{Alberto Cabezas, Adrien Corenflos, Junpeng Lao, and Rémi Louf}

\ShortHeadings{BlackJAX}{Cabezas, Corenflos, Lao, and Louf}
\firstpageno{1}

\begin{document}

\title{BlackJAX: Composable Bayesian inference in JAX}

\author{\\
\email Lead Authors \\
\name Alberto Cabezas\thanks{Department of Mathematics and Statistics, Lancaster University, UK} \email a.cabezasgonzalez@lancaster.ac.uk \\
\name Adrien Corenflos\thanks{Department of Electrical Engineering and Automation, Aalto University, Finland} \email adrien.corenflos@aalto.fi \\
\name Junpeng Lao\thanks{Google, Switzerland} \email junpenglao@google.com \\
\name Rémi Louf\thanks{.txt, France} \email remi@dottxt.co \\\\
\email Technical Contributors \\
\name Antoine~Carnec, Kaustubh~Chaudhari, Reuben~Cohn-Gordon, \\
Jeremie~Coullon, Wei~Deng, Sam~Duffield, Gerardo~Durán-Martín, \\
Marcin~Elantkowski, Dan~Foreman-Mackey, Michele~Gregori, Carlos~Iguaran, \\
Ravin~Kumar, Martin~Lysy, Kevin~Murphy, Juan~Camilo~Orduz,\\
Karm~Patel, Xi~Wang, Rob~Zinkov
\\\\
\email *Authors are ordered alphabetically within groups.
}

\editor{My editor}

\maketitle

\begin{abstract}%
BlackJAX is a library implementing sampling and variational inference algorithms commonly used in Bayesian computation. It is designed for ease of use, speed, and modularity by taking a functional approach to the algorithm's implementation. Designed from basic components to specific iterative procedures, BlackJAX allows the end user to build and experiment with new algorithms by composition. BlackJAX is written in Python using JAX to compile and run NumpPy-like samplers and variational methods on CPUs, GPUs and TPUs. The library integrates well with probabilistic programming languages by working directly with the (un-normalized) target log density function. The library is intended as a collection of low-level, composable implementations of basic statistical `atoms' that can be combined to perform well-defined Bayesian inference. It is designed for users who need cutting-edge methods, researchers who want to create complex sampling methods, and people who want to learn how these work.
\end{abstract}


\newpage
\section{Introduction}

Sampling from a probability distribution, either manually defined or constructed using probabilistic programming languages (PPLs), is a recurring topic in statistics and machine learning.
Automatic sampling software has historically been limited to Gibbs-type methods~\citep{meyer2000bugs, lunn2000winbugs, Depaoli2016JAGS}, requiring knowledge of the model structure. Black-box samplers, typically relying on Hamiltonian Monte Carlo~\citep[HMC,][]{duane1987hybrid}, allowed general, model-agnostic improvement in the applicability of the method. This was spearheaded by Stan~\citep{carpenter2017stan}, which leveraged the development of automatic differentiation. The same developments have allowed for automatically learning rich approximations via variational inference~\citep[VI,][]{jordan1999introduction} to the models of interest. Put together, these led to the creation of an array of modern PPLs that have pushed the boundaries on the feasibility of Bayesian computation~\citep{abril2023pymc, bingham2019pyro, phan2019composable}. 
While black-box samplers have paved the way, we believe that inference in today's models increasingly requires reintroducing structure-aware algorithms. 

To achieve this, BlackJAX provides users with composable inferential building blocks written using JAX~\citep{jax2018github}, such as Metropolis--Hastings~\citep{metropolis1953equation,hastings1970mcmc,robert2016metropolishastings} accept/reject step, Hamiltonian or Langevin~\citep{besag1994comments} dynamics, stochastic gradient utilities, resampling and tempering mechanisms for use within sequential Monte Carlo~\citep[SMC,][]{delmoral2006smc}, or mean field approximations \citep{jordan1999introduction}, as well as other mechanisms. 
These components are unified under a convenient, functionally-oriented API that can be combined to form new or existing algorithms immediately applicable to sequential and parallel modern computer architectures.

\section{Design principles}

BlackJAX supports sampling algorithms such as MCMC, SMC, and Stochastic Gradient MCMC (SGMCMC) and approximate inference algorithms such as VI. In all cases, BlackJAX takes a Markovian approach, whereby all the information to obtain the next iteration of an algorithm is contained in its current state. This naturally results in a functionally pure~\citep{Lonsdorf2020pure} structure, where no side-effects are allowed, simplifying parallelisation. For efficiency, auxiliary information may be included in the state too.

Once the sampling algorithm has been chosen (or designed), it is instantiated on the target function, often given by its log density. 
The sampling (or VI calibration) procedure is then carried out as a loop, updating the previous state into the current, which can be collected to compute statistics.
Additional information (for example, the acceptance probability of the proposed MCMC state) is also returned for debugging or diagnostic purposes.

\begin{lstlisting}[language=Python]
# Generic sampling algorithm:
sampling_algorithm = blackjax.sampling_algorithm(logdensity_fn, ...)
state = sampling_algorithm.init(initial_position)
new_state, info = sampling_algorithm.step(rng_key, state)

# Generic approximate inference algorithm:
approx_algorithm = blackjax.approx_algorithm(logdensity_fn, optimizer, ...)
state = approx_algorithm.init(initial_position)
new_state, info = approx_algorithm.step(rng_key, state)
position_samples = approx_algorithm.sample(rng_key, state, num_samples)
\end{lstlisting}

\subsection{Lower-level API}

Users might need a tailored algorithm for the model they are trying to sample or approximate from, or they might try out different Markov transition kernels running in parallel with various particles in an SMC algorithm, or they might use optimization to approximate the hyperparameters of the models while sampling from the rest. For any of these cases, 
BlackJAX provides access to a lower-level API, giving access to functions implementing methods with more parameters. This allows for implementing more complex methods than those available at the top level of the library.
The user-facing interface then resembles (using MCMC for illustration, but it can be any algorithm type):

\begin{lstlisting}[language=Python]
# Lower-level sampling algorithm:
init = blackjax.mcmc.sampling_algorithm.init
state = init(initial_position, logdensity_fn)
kernel = blackjax.mcmc.sampling_algorithm.build_kernel(...)
new_state, info = kernel(rng_key, state, logdensity_fn, ...)

# Lower-level approximate inference algorithm:
init = blackjax.vi.approx_algorithm.init
state = init(initial_position, optimizer, ...)
step = blackjax.vi.approx_algorithm.step
new_state, info = step(rng_key, state, logdensity_fn, optimizer, ...)
sample = blackjax.vi.approx_algorithm.sample
position_samples = sample(rng_key, state, num_samples)
\end{lstlisting}

The functional design of the library, where programs are constructed by applying and composing functions, allows the end user to build and experiment with new algorithms by applying the same mathematical logic used to design them. 
This is used, for instance, to apply different base kernels in an SMC setting or combine optimization within sampling algorithms. For a detailed example, see the implementation of the \href{https://github.com/blackjax-devs/blackjax/blob/main/blackjax/adaptation/window_adaptation.py}{window\_adaptation} scheme for adaptation of the step-size and mass-matrix appearing in HMC~\citep{betancourt2016identifying}.

\subsection{Basic components and the compositional paradigm}

All inference algorithms are composed of basic components which provide the lowest level of algorithm abstraction and are available to the user. 
With BlackJAX, researchers and practitioners can leverage all provided basic components already implemented to construct their method. 
For instance, BlackJAX contains two variants of the MH accept/reject step, starting from the computation of the accept/reject probability: 
the simpler \href{https://github.com/blackjax-devs/blackjax/blob/main/blackjax/mcmc/proposal.py#L45}{\texttt{safe\_energy\_diff}} if the proposal transition kernel is symmetric and the more general \href{https://github.com/blackjax-devs/blackjax/blob/main/blackjax/mcmc/proposal.py#L186}{\texttt{compute\_asymmetric\_acceptance\_ratio}} if the proposal transition kernel is asymmetric. Hence, the HMC algorithm uses the former while the Metropolis adjusted Langevin algorithm \citep[MALA][]{besag1994comments} uses the latter. 
After the acceptance probability of the proposal is computed, the proposal is accepted or rejected using \href{https://github.com/blackjax-devs/blackjax/blob/main/blackjax/mcmc/proposal.py#L216}{\texttt{static\_binomial\_sampling}}. In BlackJAX, this staple of MCMC can be immediately swapped for the non-reversible slice sampling algorithm of~\cite{neal2020non} simply by replacing \href{https://github.com/blackjax-devs/blackjax/blob/main/blackjax/mcmc/proposal.py#L216}{\texttt{static\_binomial\_sampling}} with \href{https://github.com/blackjax-devs/blackjax/blob/main/blackjax/mcmc/proposal.py#L246}{\texttt{nonreversible\_slice\_sampling}}.

Because all algorithms in BlackJAX are implemented as composites of these public low-level components, the end-user can then build custom algorithms that are not direct combinations of already ``well-formed'' methods such as HMC. For instance, one can use leapfrog dynamics independently of the acceptance step, giving the user more flexibility in terms of their end algorithm.

\subsection{Existing sampling libraries}
BlackJAX is the only Python library specifically aimed at users who want to use but also develop inference methods. 
In Python, \href{https://github.com/aesara-devs/aemcmc}{AeMCMC} automatically constructs MCMC samplers for probabilistic models by exploiting the symbolic graphs structure of programs written in \href{https://github.com/aesara-devs/aesara}{Aesara}\footnote{The project is on indefinite hiatus as we write this article.}. 
In Julia, \href{https://github.com/brian-j-smith/Mamba.jl}{Mamba.jl} provides a platform for implementing and applying MCMC methods to perform Bayesian analysis. Other libraries implement domain-specific algorithms, such as EMCEE~\citep{foreman2019emcee}, Dynesty~\citep{speagle2020dynesty}, and pocoMC~\citep{karamanis2022pocomc}, or are directly tied to a PPL~~\citep{bingham2019pyro,tran2019bayesian,pymc2023,carpenter2017stan,lao2020tfpmcmc}.

\section{Past impact of BlackJAX on the practice of Bayesian inference}

BlackJAX joins the already existing rich ecosystem of JAX-powered scientific libraries~\citep{jaxmd2020,wilkinson2023bayes,bonnet2023jumanji} and is directly compatible with several of them either as a client: consuming outputs from these~\citep{Pinder2022,deepmind2020jax}, or as a component, used within the libraries~\citep{phan2019composable,arviz_2019}.

BlackJAX contains a comprehensive implementation of state-of-the-art HMC algorithms, including vanilla HMC with various integrators, the no-U-turn sampling (NUTS) to choose the number of integration steps at each iteration dynamically~\citep{hoffman2014no}, and the generalized HMC algorithm~\citep{horowitz1991generalized}. It also contains adaptation schemes for the algorithms' hyper-parameters: window adaptation\footnote{\url{https://mc-stan.org/docs/reference-manual/hmc-algorithm-parameters.html}} and sophisticated calibration methods such as~\citet{hoffman2021adaptive, hoffman2022tuning}.
Consequently, several papers have leveraged BlackJAX to conduct research in various fields~\citep[see, e.g.,][]{galan2022using, price2024data, balkenhol2024candl}. Moreover, BlackJAX has made contributions to the methodological development of Bayesian inference: it has been adopted in a range of papers to develop new Bayesian sampling methods~\citep{staber2022benchmarking, cabezas2023transport, cooper2023bayesian}.

Beyond research publications, BlackJAX has found a place in courses and tutorials, for example by Darren Wilkinson\footnote{\href{functional programming for scalable statistical computing and machine learning}{\url{https://github.com/darrenjw/fp-ssc-course}}} and its use\footnote{\url{https://github.com/probml/pyprobml}} in Kevin Murphy's authoritative manuscript~\citep{pml1Book}. This usage attests to the library's recognition as a practical resource for teaching Bayesian concepts, making it accessible to a broader audience of learners and practitioners.

\section{The future of BlackJAX}

\paragraph{Enhanced portfolio.} We first aim to diversify BlackJAX's Bayesian computation methods, particularly ``meta-algorithms'' which consume base MCMC and VI methods to produce enhanced samplers. New additions include parallel and sequential tempering~\citep{marinari1992simulated,geyer1995annealing,syed2022non}; debiasing methods~\citep{jacob2020unbiased} which allow for parallelising MCMC samplers; structured VI techniques, like the Integrated Nested Laplace Approximation~\citep[INLA,][]{rue2009approximate}. Alongside these, BlackJAX will introduce more performance diagnostic tools.

\paragraph{Documentation and tutorials.} These will be expanded to cater to users at all levels, enhancing BlackJAX's usability and educational value. We plan to develop a comprehensive suite of tutorials and documentation (a ``sampling book'') detailing the integration of BlackJAX with popular probabilistic programming languages, streamlining the development process for complex probabilistic models. Additionally, the introduction of an inference database feature similar to Stan's \href{https://github.com/stan-dev/posteriordb}{posteriordb} will allow users to access a collection of posteriors for testing and benchmarking. This feature aligns with the principles of transparency and reproducibility in Bayesian inference. It will include reference implementations in probabilistic programming languages and reference posterior inferences in the form of posterior samples.

\section{Project openness and development}

BlackJAX’s source code is available under the Apache License v2.0, allowing commercial use. BlackJAX is hosted at \url{https://github.com/blackjax-devs/blackjax}. 
Anyone is welcome to contribute to the BlackJAX project. Contributions can be made in code, documentation, expert reviews of open pull requests, or other forms of support, such as case studies using BlackJAX. 
BlackJAX follows the \textbf{self-appointing council or board} open-source governance model\footnote{\url{https://www.redhat.com/en/blog/understanding-open-source-governance-models}}, that is, contributions are reviewed by core contributors, and breaking decisions about the BlackJAX project are made by consensus among these.
A comprehensive test suite is run automatically by a continuous integration service before code can be merged into the code base. As of the writing of this article, the test coverage of the library is 99\%.

\vskip 0.2in
\bibliography{jmlr_oss}

\end{document}